 \documentclass[onecolumn]{pasj00}
\usepackage{natbib}

\SetRunningHead{Takei et al.}{WHIM around A2218}
\Received{2006/07/19}
\Accepted{2006/08/01}

\begin{document}

\title{
Search for Oxygen Emission from 
Warm-Hot Intergalactic Medium around 
A2218
with {\it Suzaku}
}

\author{
Yoh \textsc{Takei}\altaffilmark{1},
Takaya \textsc{Ohashi}\altaffilmark{2},
J. Patrick \textsc{Henry}\altaffilmark{3},
Kazuhisa \textsc{Mitsuda}\altaffilmark{1}, \\
Ryuichi \textsc{Fujimoto}\altaffilmark{1}, 
Takayuki \textsc{Tamura}\altaffilmark{1},
Noriko Y. \textsc{Yamasaki}\altaffilmark{1}, \\
Kiyoshi \textsc{Hayashida}\altaffilmark{4}, 
Noriaki \textsc{Tawa}\altaffilmark{4}, 
Kyoko \textsc{Matsushita}\altaffilmark{5},
Mark W. \textsc{Bautz}\altaffilmark{6},
John P. \textsc{Hughes}\altaffilmark{7},\\
Grzegorz M.  \textsc{Madejski}\altaffilmark{8},
Richard L. \textsc{Kelley}\altaffilmark{9},
Keith A. \textsc{Arnaud\altaffilmark{9}}
}
\altaffiltext{1}{
Institute of Space and Astronautical Science,
Japan Aerospace Exploration Agency,\\
3-1-1 Yoshinodai, Sagamihara, Kanagawa 229-8510
}
\email{takei@astro.isas.jaxa.jp}
\altaffiltext{2}{Department of Physics, Tokyo Metropolitan University,
1-1 Minami-Osawa, Hachioji, Tokyo 192-0397}
\altaffiltext{3}{Institute for Astronomy, University of Hawai'i, 2680 Woodlawn Drive, Honolulu, HI 96822, USA}
\altaffiltext{4}{Department of Astrophysics, Faculty of Science, Osaka University, Toyonaka 560-0043}
\altaffiltext{5}{Department of Physics, Tokyo University of Science, 1-3 Kagurazaka, Shinjuku, Tokyo 162-8601}
\altaffiltext{6}{Kavli Institute for Astrophysics and Space Research, 
Massachusetts Institute of Technology,\\ 
70 Vassar Street, Building 37, Cambridge, MA 02139, USA
}
\altaffiltext{7}{Department of Physics and Astronomy, Rutgers University, Piscataway, NJ 08854-8019, USA}
\altaffiltext{8}{Stanford Linear Accelerator Center, 2575 Sand Hill Road, Menlo Park, CA 94025, USA}
\altaffiltext{9}{NASA Goddard Space Flight Center, Code 662, Greenbelt, MD 20771, USA}


\maketitle

\begin{abstract}
We searched for redshifted O emission lines from the possible warm-hot
intergalactic medium (WHIM) surrounding the cluster of galaxies A2218
at $z=0.1756$ using the XIS instrument on {\it Suzaku}. This cluster
is thought to have an elongated structure along the line of sight
based on previous studies. We studied systematic uncertainties in
the spectrum of the Galactic emission and in the soft X-ray response
of the detectors due to the contamination building up on the XIS filters.
We detected no significant redshifted O lines, and set a tight
constraint on the intensity with upper limits for the surface
brightness of O\emissiontype{VII} and O\emissiontype{VIII} lines of
$1.1\times10^{-7}$ and
$3.0\times10^{-7}$~$\mathrm{photons~cm^{-2}~s^{-1}~arcmin^{-2}}$,
respectively.  These upper limits are significantly lower than the
previously reported fluxes from the WHIM around other clusters of
galaxies.  We also discuss the prospect for the detection of
the WHIM lines with {\it Suzaku} XIS in the future.

\end{abstract}

\section{Introduction}
Based on several N-body simulations of cosmological
large-scale structure formation
\citep[e.g.,][]{cen99:_where_are_baryon,
dave01:_baryon_warm_hot_inter_medium,2003ApJ...594...42C}, a
significant (30--50\%) fraction of baryons is thought to reside in the
form of gas in a `warm-hot' phase ($T=10^{5-7}$~K), which is hard to
detect with the instruments currently in operation.  This warm-hot
gas, whose density is $10^{-6}$--$10^{-4}\mathrm{~cm^{-3}}$, is called
the warm-hot intergalactic medium (WHIM)\@.  The firm detection of the
WHIM is important because it is the most promising candidate for the
``missing baryons''; i.e., it is thought to explain the discrepancy
between the baryon density observed in the local universe
\citep{fukugita98:_cosmic_baryon_budget} and that in the
distant universe
\citep{rauch97:_opacit_ly_fores_implic_omega} or that calculated from
the observed fluctuations of the cosmic microwave background
\citep{spergel03:_first_year_wilkin_microw_anisot}.

The WHIM may be detected via emission or absorption lines from highly
ionized elements in UV or X-ray spectra.  
Given the relative elemental abundances,
--- on the order of 0.1 to 0.3 of Solar ---
oxygen is the most promising element to
provide detectable transitions at the expected temperatures.
Over 40 systems show
O\emissiontype{VI} absorption features, based on observations from
{\it FUSE} and {\it HST} \citep[e.g.][]{2005ApJ...624..555D}.
However, their analysis shows that the baryon density in the
temperature range $T=10^{5-6}$~K is about an order of magnitude lower
than the expected WHIM level. Therefore, most of the missing baryons
seem to reside in a hotter phase at $T=10^{6-7}$~K, which can be
probed with X-rays.  
There is only one
sightline with a likely detection of the WHIM via 
O\emissiontype{VII} or O\emissiontype{VIII} absorption lines so far
\citep{2005ApJ...629..700N}.
Their results suggest that most of the
missing baryons can be explained by
the absorbing clouds inferred from their data.
However, recent work by 
\cite{2006astro.ph..4515R} and \cite{2006astro.ph..4519K}
showed that the detection with the Chandra LETGS by
\cite{2005ApJ...629..700N} was not consistent with RGS observations 
of the same sight line.  Thus,
the existence of the WHIM in the hotter phase is not yet confirmed.

Several possible detections of WHIM emission have been reported.
\citet{2003A&A...397..445K} and \citet{2004JKAS...37..375K}
report the detection of redshifted
O\emissiontype{VII} and O\emissiontype{VIII}
emission lines as well as soft X-ray
($E<0.5~\mathrm{keV}$) excesses in the spectra of 
7 clusters of galaxies out of 21 measured systems.
\citet{finoguenov03:_xmm_newton_x_coma} also reported detection of
O\emissiontype{VII} and O\emissiontype{VIII} emission lines in the
outskirts of the Coma cluster.
\citet{fujimoto04:_probin_warm_hot_inter_medium} and Takei et al.\
(2006) report not only emission lines in the spectrum of cluster
outskirts but also absorption lines in the spectrum of a quasar behind
the clusters, although the significance of the absorption measurement
is not high ($<3\sigma$).
\citet{fujimoto04:_probin_warm_hot_inter_medium} observed
O\emissiontype{VIII} features in the Virgo cluster, while Takei et
al.\ (2006) detected Ne\emissiontype{IX} in the Coma cluster.  They
estimated the path length of absorbing/emitting medium assuming that the
absorber and the emitter are the same cloud, and concluded 
that the length
scale of the medium exceeds the virial scale of the clusters.  
There are counterarguments against the discovery of
O\emissiontype{VII} or O\emissiontype{VIII} emission lines
associated with clusters.
For example, \citet{2005A&A...443...29B} and 
\citet{2006ApJ...644..167B} concluded that some of the emission
lines observed by \citet{2003A&A...397..445K} are likely
due to field-to-field variation of the soft X-ray background
or Galactic emission.
Such controversies occurred because the CCD instruments cannot 
clearly distinguish
O lines emitted by  the gas
associated with clusters of moderate redshift ($z\lesssim 0.05$)
from those of Galactic origin.
These arguments suggest the importance of careful target selection
and accurate measurement of background level.
Note that
the observations of redshifted emission lines
also give important clues to solve
another mystery
``cluster soft excess'', the excess emission in $E<0.3~\mathrm{keV}$,
reported since the 1990s in the extreme ultraviolet and the X-ray
bands from 
{\it EUVE} and {\it ROSAT}
\citep[e.g.,][]{lieu96:_diffus_coma, 1996ApJ...458L...5L,
1998ApJ...498L..17M,2004ApJ...605..168B}.  
Thus, the study of X-ray emission in cluster
outskirts brings us important information about the cluster
system itself and its environment.

%%%
Since there are only a few reports of possible emission lines from
the WHIM, further sensitive study of other systems is desirable.  In
this paper, we report our observation of A2218 with the XIS
(Koyama et al.\ 2006) onboard {\it Suzaku} (Mitsuda et al. 2006).
A2218 is a well-known cluster because of its strong gravitational
lensing arcs.  Its redshift is 0.1756 
\citep{struble99:_compil_redsh_veloc}, where one
arcminute corresponds to 179 kpc with the cosmology we have adopted.
The cluster is thought to be still in a dynamically young state --- X-ray
analysis suggests that the cluster core is not in hydrostatic
equilibrium due to an ongoing or recent merger
\citep{2005A&A...433..777P, 2002ApJ...567..188M} in the line-of-sight
direction. The distribution of galaxy velocities based on the optical
study of \citet{1997ApJ...490...56G} indicates substructure in the line
of sight, and \citet{2005MNRAS.359..417S} studied the mass structure in
detail and have concluded that A2218 is unrelaxed.  The dynamics of
the cluster is considered to cause the discrepancy between the mass
estimated from gravitational lensing and that determined from
X-ray observations; gravitational lensing indicates $\sim 3$ times higher
mass than X-ray estimation for the central ($\sim20''$) region.
These observations indicate an elongated structure in the
line-of-sight direction and suggests existence of a large-scale
filament in this direction.  If this is the case, we would expect a
high surface brightness of the WHIM emission. The fairly large
redshift of A2218 would also help us observe a significant energy
shift of emission lines with the XIS instrument. Thus, A2218 seems to
be one of the most suitable targets to search for the WHIM emission
from Suzaku.

We assume a Hubble constant of 70 km s$^{-1}$ Mpc$^{-1}$ or $h_{70}=1$
and $\Omega_\mathrm{m} = 0.3$, $\Omega_\mathrm{\Lambda} = 0.7$.
The solar metal abundances used in this paper are given by
\cite{anders89:_abund}.
Unless otherwise
stated, errors in the figures are at the 68~\% confidence level and at
the 90~\% confidence level in the text and tables.

%%%%%%%%
\section{Observations}

\begin{figure}[hbt]
\begin{center}
\FigureFile(0.7\columnwidth,clip){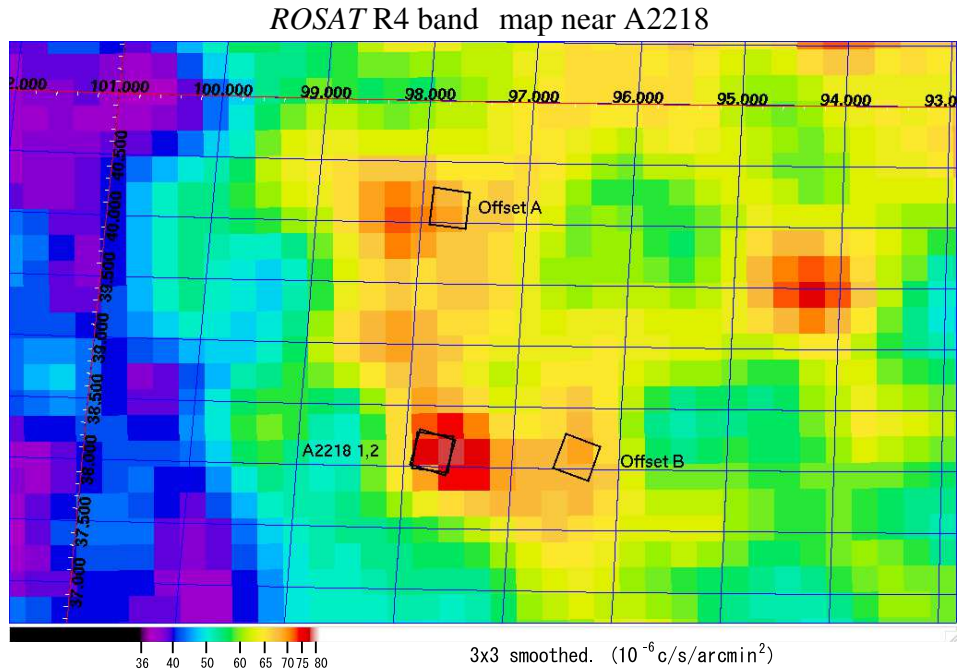}
\end{center}
\caption{
{\it ROSAT} R4 band image around A2218
\citep{1997ApJ...485..125S} in Galactic coordinates.  
The field of view of XIS in the four observations are indicated by black
squares.
Most of the emission in
this band is O\emissiontype{VII} and O\emissiontype{VIII} line.
}
\label{fig:a2218rosatmap}
\end{figure}

We carried out four observations to study the warm-hot gas in
the A2218 vicinity with the XIS instrument onboard {\it Suzaku}: two
on the cluster and two in offset regions whose locations are different
from each other.  The offset observations were performed to measure
the foreground Galactic contribution, which can produce a large
ambiguity in the study of soft emission around clusters.
Fig.~\ref{fig:a2218rosatmap} is a map of the {\it ROSAT} R4 band after
removing bright point sources \citep{1997ApJ...485..125S}.  This
energy band is sensitive to O\emissiontype{VII} and
O\emissiontype{VIII} emission lines, which are strong in the Galactic
emission, and hence is suitable to study their fluctuations.

We examined the {\it ROSAT} map (Fig.~\ref{fig:a2218rosatmap}) and
selected the location of offset pointings so that the Galactic
emission level is similar to that in the vicinity
of A2218. Also, the distance
from A2218 is large ($\gtrsim 1^\circ$ or 10.7 Mpc at the A2218
redshift), so the data from the offset regions should be free from the
putative large-scale filament around A2218.  The observed locations
are shown in Fig.~\ref{fig:a2218rosatmap} and
Table~\ref{tab:a2218position}, as well as the position of A2218.
Hereafter we call the four observations A2218-1, A2218-2, Offset-A,
and Offset-B, respectively.  A2218-1 and Offset-A were observed in
early October, 2005, and the others in late October. The observation
sequence numbers and dates are summarized in
Table~\ref{tab:a2218observation}. 

The XIS is an X-ray CCD camera, which consists of three
front-illuminated (FI) sensors and one back-illuminated (BI) sensor.
The camera covers the 0.2--12~keV energy band with an energy resolution of
130~eV at 5.9~keV and 40~eV at 0.5~keV.  Its field of view is a square
of $18'\times18'$.  The BI and FI sensors have different
advantages:
the former has higher quantum efficiency in the soft energy band,
while the latter shows a lower internal background level.  An
important characteristic of the XIS is the absence of a large
low-energy tail in the pulse-height distribution function, even in the
very soft energy ($E\lesssim 0.5~\mathrm{keV}$) band.  Those
characteristics give the XIS an excellent capability to study the
low-energy emission lines, such as from the WHIM.  After the launch of
{\it Suzaku}, a time and position dependent contamination of the XIS
optical blocking filter (OBF) was found.  The source is probably 
outgassing from the satellite. The level of contamination increases with
time and is different from sensor to sensor.  The time and position
dependence of the contamination thickness has been empirically modeled
by the XIS team.  At the time of our observations, the effective area
at 0.5~keV was on average 25\% lower than that of the pre-launch
calibration.
See Koyama et al. (2006) for a fuller description of
the instrument.
The two observations of A2218 were 25 days apart, and the count rates
at 0.5 keV were different by 9\% and 13\% for the BI and the combined FI
sensors, respectively. This is consistent with the expected drop of 8\%
for both sensors within statistical errors.

\begin{figure}[hbt]
\begin{center}
 \FigureFile(0.49\columnwidth,clip){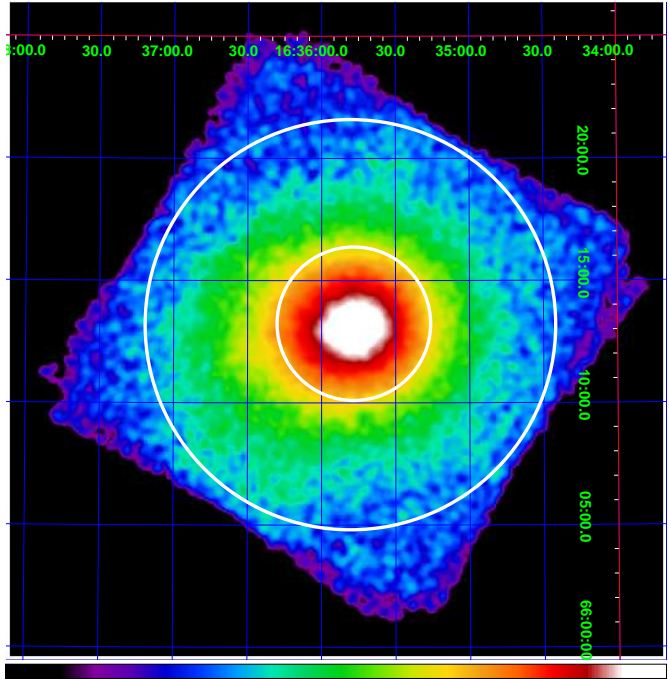}
\end{center}
\caption{Image of A2218 in the 1.0--4.0 keV band
observed with the XIS.  The events of the two
observations obtained by four CCDs were summed.
The spectra were extracted from the annulus between the two white circles.
Vignetting effects were not taken into account for this image and 
background events were not subtracted.
The grid indicates the coordinates in J2000.0.
}
\label{fig:a2218image}
\end{figure}

The XIS instrument was operated in normal mode in all the
observations presented in this paper.  
We used event files of version 0.7 products\footnote{
Version 0 processing is an internal processing applied to the
Suzaku data obtained during the SWG phase.
Aspect correction and fine tuning of the event time tagging are
skipped in this version.  Times when the elevations from the
bright and dark
Earth are $>20^{\circ}$ and  $>5^{\circ}$ are excised in standard
data processing.
}.
Events of $3\times3$ and $5\times5$ observation modes
were combined.  
The first few kiloseconds of each observation, when
the pointing direction had not yet stabilized, were excluded.  The
resulting image of A2218 in the 1.0--4.0~keV energy band is shown in
Fig.~\ref{fig:a2218image}. Data from the two observations and the
four CCDs are combined.  The white circles indicate the region where
we extracted the spectra. The radii are $3'$ and $8'$ from the cluster
center, corresponding to 540 and 1430 kpc at the source, respectively.

\begin{table}[hbt]
\begin{center}
\caption{Position of the A2218 and offset pointing
\label{tab:a2218position}
}
{ \small
\begin{tabular}{lcc}
\hline \hline
& (RA, Dec) & ($l$, $b$) \\ \hline
A2218 &
(\timeform{16h35m54s}, \timeform{66D13'00''}) & (97.7449, 38.1235)\\
Offset-A &
(\timeform{16h17m48s}, \timeform{65D27'36''}) & (97.7423, 40.1239)\\
Offset-B &
(\timeform{16h39m31s}, \timeform{66D13'31''}) & (96.4000, 38.1002)\\
\hline
\end{tabular}
}
\end{center}
\end{table}

\begin{table}[hbt]
\begin{center}
\caption{{\it Suzaku} observations of A2218 and offset pointing
\label{tab:a2218observation}
}
{%\scriptsize
\begin{tabular}{lccccc}
\hline \hline
& Sequence & Date & Exposure & Net exposure& Net exposure \\ 
& number & & after processing
& using all COR & using COR$>8~\mathrm{Gev}~c^{-1}$ \\ \hline
A2218-1 & 100030010 & 2005-10-01--2005-10-02 & 46.4~ks & 44.9 ks & 38.2 ks\\
A2218-2 & 800019010 & 2005-10-26--2005-10-27 & 32.8~ks & 32.3 ks & 28.8 ks\\
Offset-A & 100030020 & 2005-10-02--2005-10-03 & 44.6~ks & 44.6 ks & 39.0 ks\\
Offset-B & 800020010 & 2005-10-27--2005-10-27 & 15.0~ks & 15.0 ks & 12.0 ks\\
\hline
\end{tabular}
}
\end{center}
\end{table}

\begin{figure}[hbt]
\begin{center}
\FigureFile(0.47\columnwidth,clip){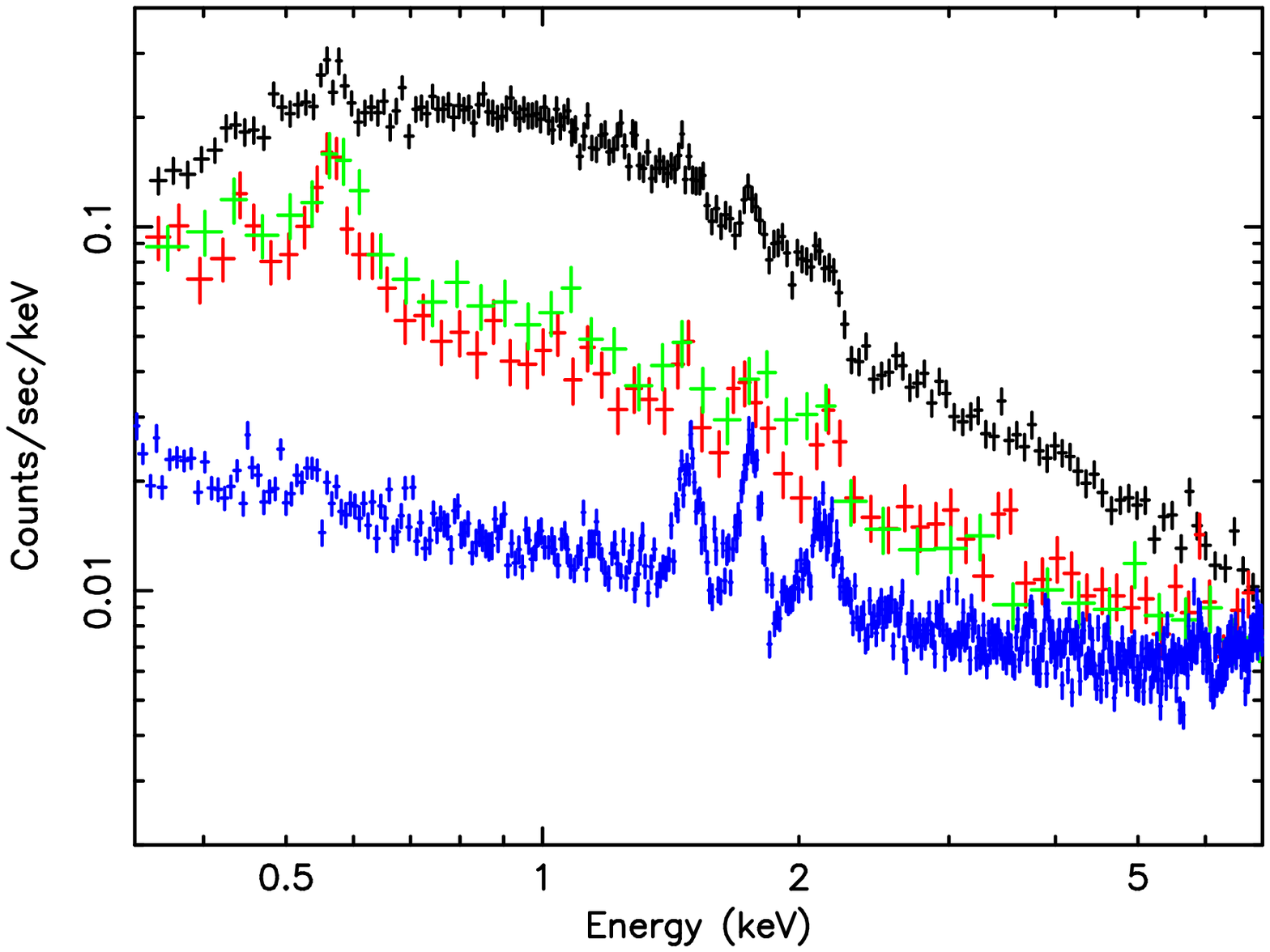}
\hspace{0.03\columnwidth}
\FigureFile(0.47\columnwidth,clip){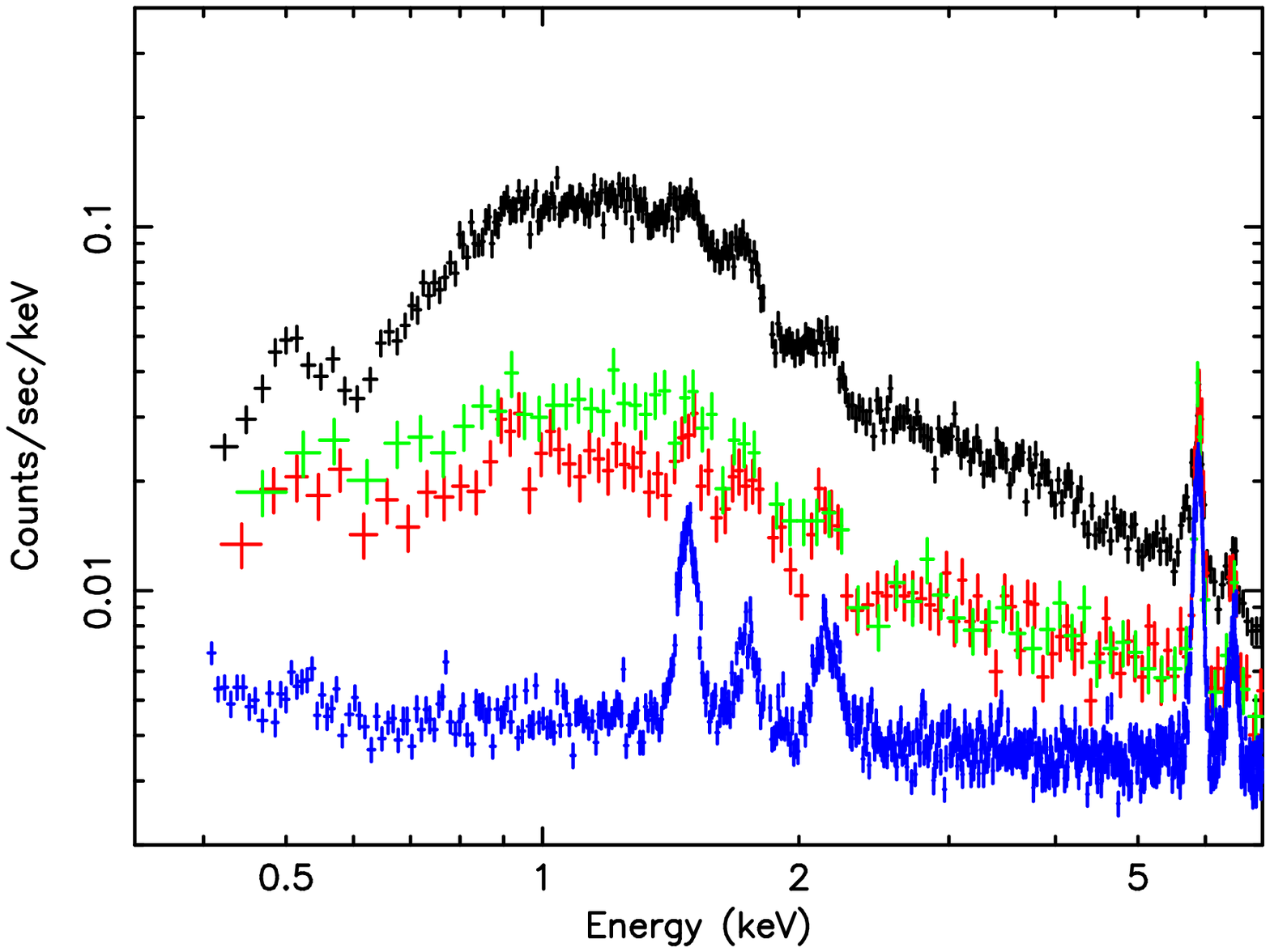}
\end{center}
\caption{
Spectra of A2218 taken with the (left) BI and (right) FI chips,
where three FI spectra were
averaged.
Black, red, green and blue points show A2218, Offset-A, Offset-B, night
earth spectra, respectively.
The strong peaks at 5.9 keV and 6.5 keV in the right panel are
due to the calibration sources.
}
\label{fig:a2218comparespectra}
\end{figure}

\section{Spectral analysis}

\subsection{Analysis method}
\label{sec:Analysis-method}

We removed flickering CCD pixels, which cause a large noise component
below 0.5~keV\@.  We also excluded apparent point sources that were
found in the Offset-A image.  
Although the fluxes of these point sources were lower than the 
detection limit for the A2218 image, we removed them because
we are interested in the upper limits of the soft excess emission.
The spectra were then extracted in the
annular region between $3'$ and $8'$ from the cluster center using
XSELECT distributed in HEASOFT 6.0.4.  We define the cluster center as
(DETX, DETY) = (499, 530) in detector coordinates for all observations
and all sensors.  Note that the boresight of all XIS sensors
coincide within $20''$.   We have chosen the inner radius to be $3'$
in order to exclude the bright central region, in which the strong
emission of the hot ICM hampers the study of warm-hot emission. We
also excluded the region outside of $8'$ because the position
dependence of the contamination on the OBF is not well known there.
Although the XIS sensors are always illuminated by $^{55}$Fe
calibration sources at the field edge, 
we did not remove these regions
in the present analysis.

To estimate and subtract the internal background, we also extracted
spectra from night Earth observations distributed by
the XIS team.  It is known that the variation of the internal
background spectrum is well correlated with the cut-off rigidity (COR)
at the position of the satellite; the
smaller the COR value is, the larger the background level becomes.  To
estimate accurately the internal background level, we collected events
when the detector was looking at the dark Earth (night Earth events)
and sorted them by COR\@. We extracted the spectrum for each 
$1~\mathrm{GeV}~c^{-1}$
interval of COR, and then added them weighted by exposure time
for the respective COR range in the actual observation.  This process
gives different night Earth spectra for the four observations, since
the detailed distribution of COR was different among the observations.
The background spectra were extracted from the same region in
detector coordinates as the corresponding observation in order to avoid
possible systematic effects due to positional variation of the
detector background.

We used response matrix files (RMFs) ae\_xi[0123]\_20060213(c).rmf
distributed by the XIS team for spectral fitting.  On the other hand,
we generated the ancillary response files (ARFs) with the arf builder
`xissimarfgen', which is based on ray-tracing (Ishisaki et al.\ 2006).
The change of the effective area with time in the soft X-ray energy
range, due to the increase of OBF contamination, was taken into
account by the following method.  The ARFs were created separately for
each observation, according to the observation date.  We used the
ae\_xi[0123]\_contami\_20060525.fits contamination tables to model the
composition and position dependence of the contaminant, in which the
C/O ratio of contaminant was assumed to be 6.  Note that the effective
area for diffuse sources depends on the sky distribution of the flux,
because the quantum efficiency varies with the detector position due
to vignetting and contamination thickness.  We adopted the Suzaku XIS
image (1.0--4.0~keV)
of A2218 for the incident flux distribution, in which the two
observations by the four sensors were all summed (see
Fig.~\ref{fig:a2218image}). We assumed a uniform flux distribution for
the offset pointings.

In the analysis shown below, the spectra and ARFs for A2218-1 and
A2218-2 were added using the ftools mathpha and addarf, respectively.
Further, the spectra and response (RMFs and ARFs) of the three FI
sensors were combined using ftools mathpha and marfrmf, respectively.
We confirmed that consistent results were obtained with this treatment
within the statistical error.
We fitted spectra of the offset observations in the 
0.35~keV $<E<$ 5.0~keV energy range for the BI and 
0.40~keV $<E<$ 5.0~keV
except the energy of anomalous response at the Si K-edge,
1.825--1.840~keV, 
for the FI.
When we fitted the spectra of A2218, the energy range was
extended up to 7.0 keV in order to cover the Fe-K lines.  We excluded
5.85--5.95~keV and 6.45--6.55~keV, because the slight difference
in intensity of Mn~K emission lines from the $^{55}$Fe calibration sources 
between A2218 observations and
night Earth observations causes relatively large
residuals.
The spectra of A2218 (black), Offset-A
(red), and Offset-B (green), as well as the night Earth spectra (blue)
are shown in Fig.~\ref{fig:a2218comparespectra}.  Left and right
panels are for BI and FI spectra, respectively.  The difference in the
detector area due to point-source exclusion was corrected in the
figure for illustrative purposes.
We mainly performed simultaneous fits for the two spectra obtained by
the BI and the combined FI sensors. However, it was also confirmed that
the spectrum with the BI sensor only gave consistent results. This
indicates that the uncertainty in the FI response around 0.5 keV is
not crucial compared with the statistical errors in the BI spectrum.

Evaluation of the systematic uncertainties is crucial in constraining
the emission from the WHIM\@.  Firstly, the emission is in the soft
X-ray region where the XIS has various systematic uncertainties, so 
we have
to look into many possible effects.  Secondly, the observation was
carried out only three months after the launch of {\it Suzaku}, and
time variation of the detector response needs to be carefully
considered.

The first uncertainty is the detector response.  The largest
systematic uncertainty in the detector response is the
thickness of contaminant on the OBF of the XIS\@.  In the 0.4--0.6 keV
band, most of the photons we observe are thermal emission from the
ICM\@.  If we underestimate the contaminant thickness, the flux from
the ICM would be overestimated in $E\lesssim 1$~keV, because the
temperature and abundance are strongly constrained by the data in
the higher energy range $E\gtrsim 1$~keV\@.  This naturally leads to an
underestimation of the soft-excess flux.  Therefore, we also
considered a {\it thicker} contaminant model at the upper limit of the
uncertainty range.  We generated ARFs with 20\% thicker contaminant
using xissimarfgen assuming the observation occurred seventeen days
later than it actually did.  
This 20\% increase, which  exceeds the
uncertainty quoted by the XIS team ($\sim 10\%$; Koyama et~al. 2006),
is a reasonable value, given  the additional uncertainties in
the response to diffuse emission and in early observations.
The energy resolution of the XIS worsens with time. At the
time of our observations, the energy resolution had degraded by
$\sim20$~eV (FWHM) at 5.9~keV (from 130~eV to 150~eV).
This effect was investigated by smoothing the response 
with a Gaussian.  Since we have no information about the degradation
at $\sim0.5$~keV,  we tried Gaussians with sigmas of
5~eV, 10~eV, 15 eV, 20 eV, 30 eV  and 35~eV. 
With this smoothing, the spectral resolution
at $\sim$0.5 keV increases from 40 eV to 42 eV, 46 eV, 53 eV,
62 eV, 81 eV and 91 eV, respectively.
The results are presented later.
The internal background of the XIS is quite stable, and there were no
``background flares'' during our observations.  However, when COR is
low (COR$\lesssim 6~\mathrm{GeV}~c^{-1}$), the internal background
level becomes comparable to the X-ray background
in the $E>3$~keV range.  
Therefore, we also extracted spectra with the condition COR
$>8~\mathrm{GeV}~c^{-1}$ to examine this systematic effect.

Another uncertainty is the spatial variation of the Galactic emission.
The {\it ROSAT} R4 map shows $\sim 10\%$ variations among the A2218,
Offset-A and Offset-B fields.  The purpose of the two offset
observations (Offset-A and Offset-B) was to look at the spectral
variation in the Galactic emission and to include it in the
analysis. Besides the inclusion of this variation, we also extended
our upper limit for the soft excess by assuming 10\% fainter Galactic
emission.

%%%%%%%%%%
\subsection{Offset pointings}

Before the analysis of the A2218 spectrum, we analyzed data from the
offset pointings to estimate the spectrum of the Galactic emission.
The diffuse X-ray background at high Galactic latitude can be divided
into three components: i.e.\ the local hot bubble (LHB), the Milky Way
halo (MWH) and the extragalactic power-law (CXB) components 
\citep{1998ApJ...493..715S}.
Typical temperatures are about 0.1 keV for the LHB and
0.2--0.3 keV for the MWH, while the CXB spectrum has a photon index of
$1.4$.  The LHB is thought to surround the solar system with a
$\sim100$~pc scale, and hence it has no Galactic absorption.  The CXB
component is extragalactic and known to be uniform over the entire
sky.  In contrast, the level of LHB and MWH components vary from
position to position.  \citet{2002A&A...389...93L} reported that the
mean deviation of the 0.2--1.0~keV intensity is $\sim35\%$ from field
to field.

Our purpose here is to constrain the spectra of the two Galactic
components so as to obtain reliable background data for the estimation
of the warm emission around A2218.  Spectra for an annular region with
$r = 3'-8'$ were produced for the Offset-A and Offset-B data.  XSPEC
v11.3.2r was used in the spectral analysis.  The spectra, from which non
X-ray background was subtracted, were fitted with a model for the sum
of LHB, MWH and CXB\@.  CXB and MWH components were absorbed by
Galactic absorption characterized by $N_{\rm H} = 3.24\times
10^{20}~\mathrm{cm}^{-2}$ \citep{dickey90}.  The collisionally ionized
thermal plasma model APEC was used in XSPEC to fit the LHB and MWH
spectra, with abundance and redshift fixed to 1 solar 
and 0,
respectively.  The photon index of the power-law CXB model and the
temperature of the LHB were fixed to 1.4 and 0.08 keV, respectively.  The
normalizations of the FI and BI sensors were allowed to take different
values.

\begin{figure}[hbt]
\begin{center}
\FigureFile(0.47\columnwidth,clip){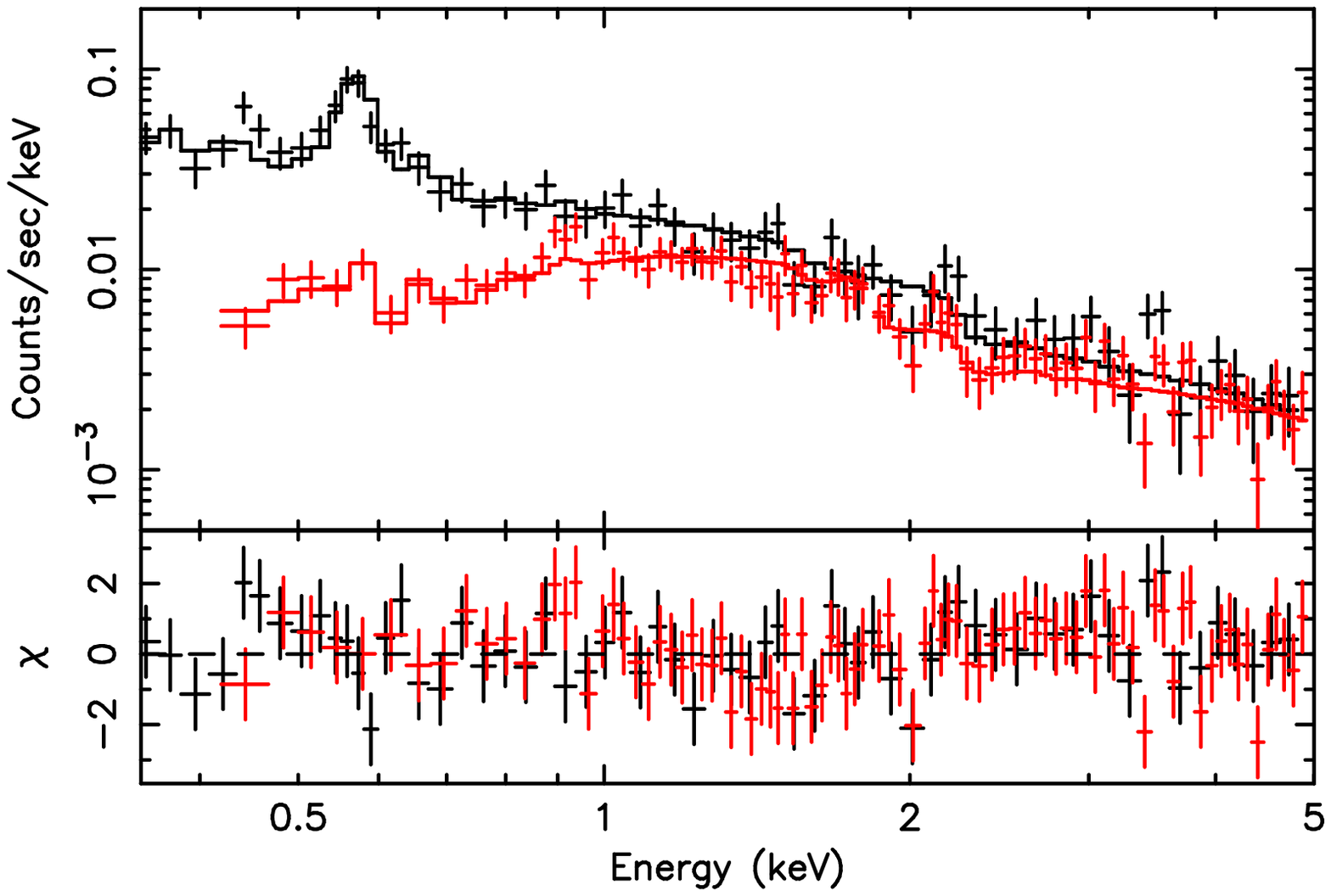}
\hspace{0.03\columnwidth}
\FigureFile(0.47\columnwidth,clip){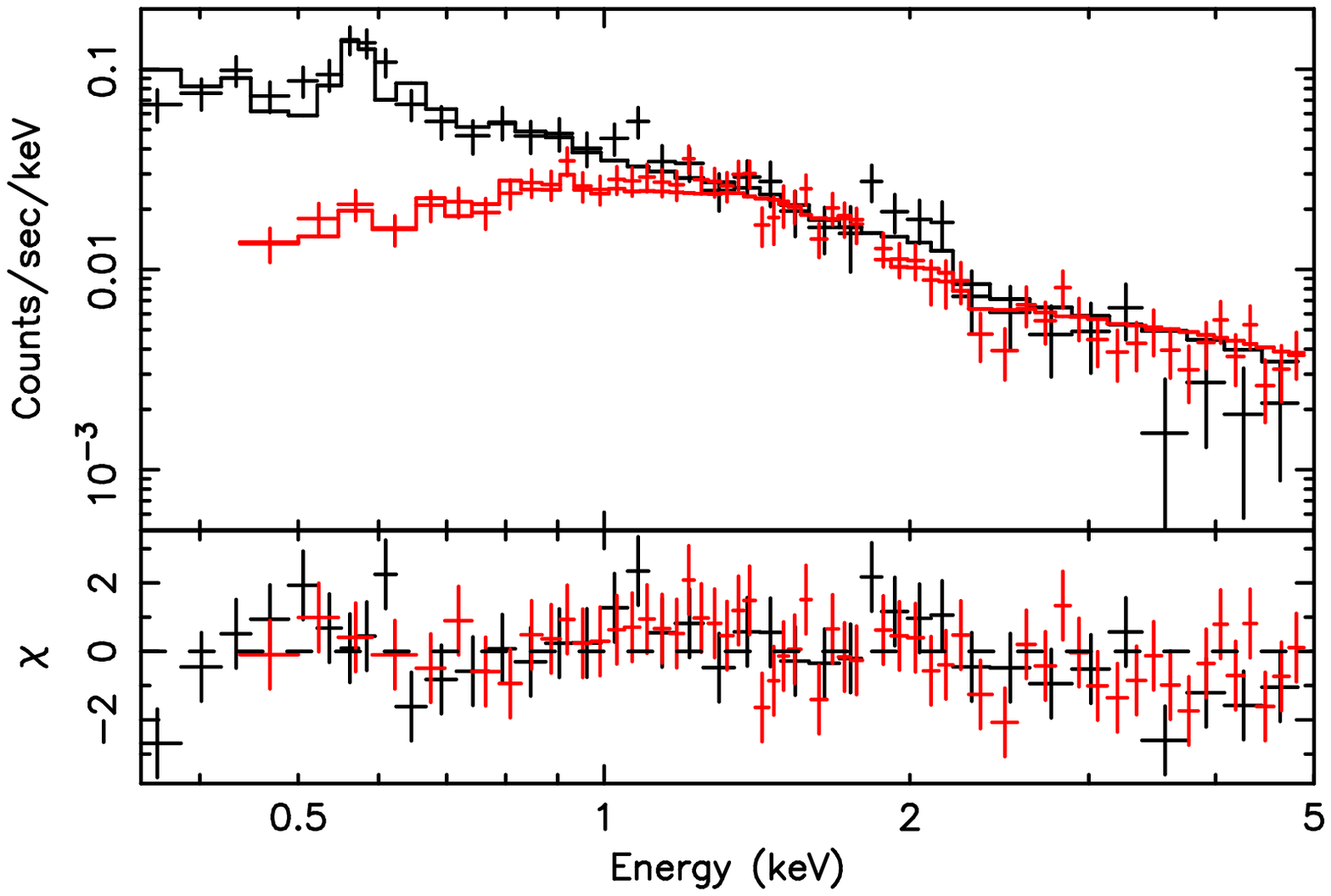}
\end{center}
\caption{
Spectra and best-fit models, and residuals 
of offset observations. (left) Offset-A, (right) Offset-B.
Black and red represents BI and FI sensors.
}
\label{fig:offset-3-8}
\end{figure}

The best-fit parameters and the model curves are shown in
Table~\ref{tab:offsetspectra} and Fig.~\ref{fig:offset-3-8}.  In
Table~\ref{tab:offsetspectra}, parameters except for the FI/BI ratio
are for the BI spectra. Acceptable fits were obtained for the two
spectra and the parameters of the Galactic emission were well
constrained. The normalization of the CXB component agrees with 
previous reports 
\citep{2002PASJ...54..327K}.
The flux in the 0.5--2.0~keV band of Offset-A and Offset-B observation
differs by $\sim10\%$.
Not only the flux, but also the shape of the best-fit model is
different between the two observations, which is either due to the
poor statistics of the Offset-B observation or real variations
in the Galactic emission.

\begin{table}[hbt]
\begin{center}
\caption{Best-fit parameters of offset pointings
\label{tab:offsetspectra}
}
{\normalsize
\begin{tabular}{lcc}
\hline \hline
Parameter & Offset-A & Offset-B \\
\hline 
LHB $kT$ (keV) & 0.08 (fixed) & 0.08 (fixed) \\
LHB Normalization$^\mathrm{a}$
 & $3.2\pm1.0\times10^{-6}$ 
 & $6.2^{+0.7}_{-1.7}\times10^{-6}$ \\
MWH $kT$ (keV) & $0.158^{+0.028}_{-0.018}$ & $0.248^{+0.033}_{-0.031}$\\ 
MWH Normalization$^\mathrm{a}$ & $6.7^{+3.3}_{-2.5}\times10^{-7}$
  & $5.0^{+1.5}_{-1.0}\times10^{-7}$\\
CXB Photon index & 1.4 (fixed) & 1.4 (fixed)\\
CXB Normalization$^\mathrm{b}$ &  $8.5\pm0.5\times10^{-7}$
 &  $8.5\pm0.7\times10^{-7}$\\
FI/BI ratio & $0.84\pm0.05$ & $1.04\pm0.09$\\
\hline
$\chi^2/\mathit{dof}$ & 154.03/149 & 100.45/93 \\
\hline
\multicolumn{3}{l}{ \parbox{10cm}
{$^\mathrm{a}$~
$\int n_\mathrm{e}n_\mathrm{H}~dV/4\pi(D_\mathrm{A}(1+z))^2$ per
solid angle in units of $10^{14}~ \mathrm{cm^{-5}~ arcmin^{-2}}$,
where $n_\mathrm{e}$ is the electron density,
$n_\mathrm{H}$ the hydrogen density, and $D_\mathrm{A}$ 
the angular size distance.
}
}\\
\multicolumn{3}{l}{$^\mathrm{b}$~
 In units of $\mathrm{photons~cm^{-2}~s^{-1}~keV^{-1}~arcmin^{-2}}$ at 1 keV}
\end{tabular}
}
\end{center}
\end{table}

\subsection{A2218}

\subsubsection{Single temperature model}
\label{sec:Single-temp-model}

\begin{figure}[hbt]
\begin{center}
\FigureFile(0.47\columnwidth,clip){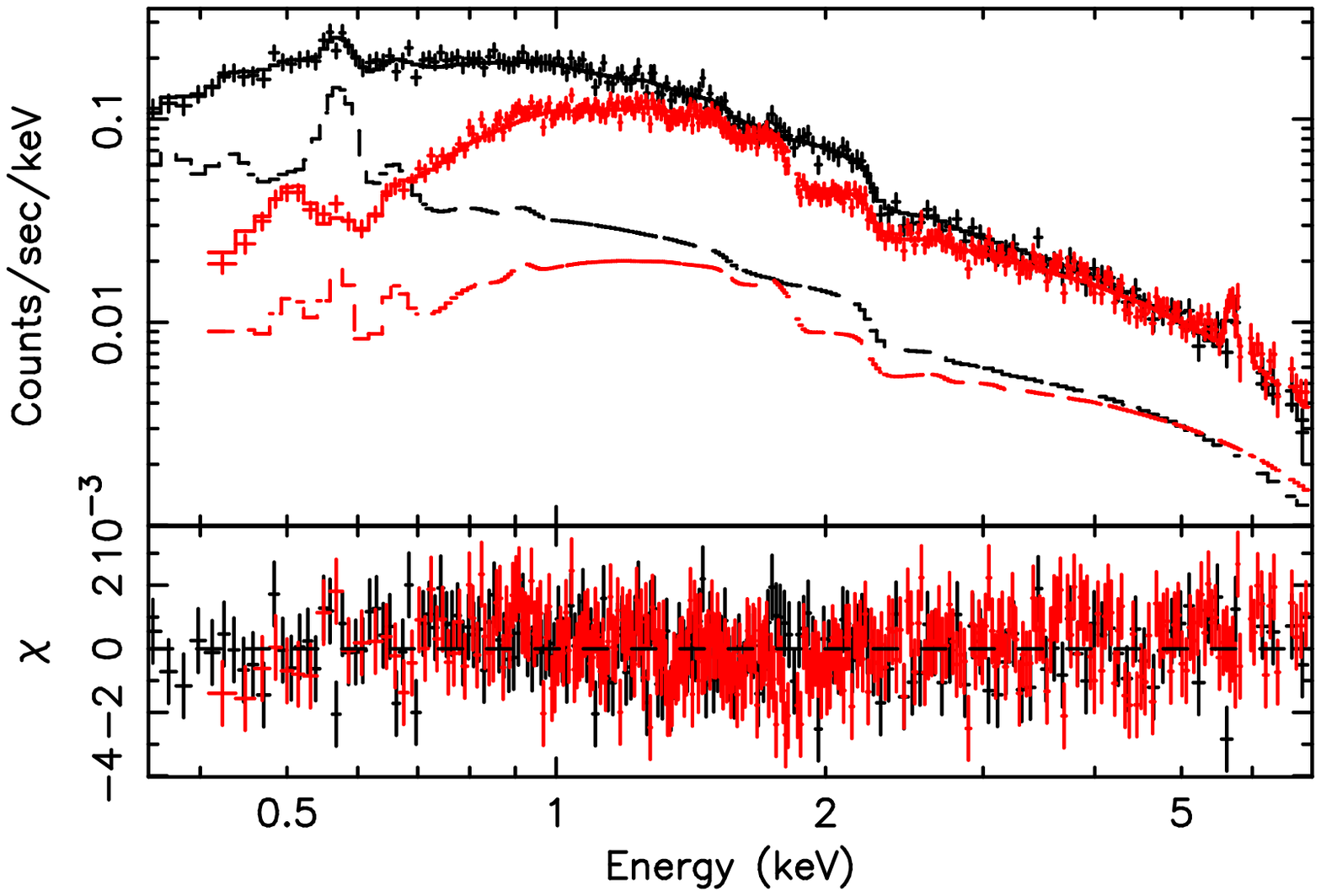}
\hspace{0.03\columnwidth}
\FigureFile(0.47\columnwidth,clip){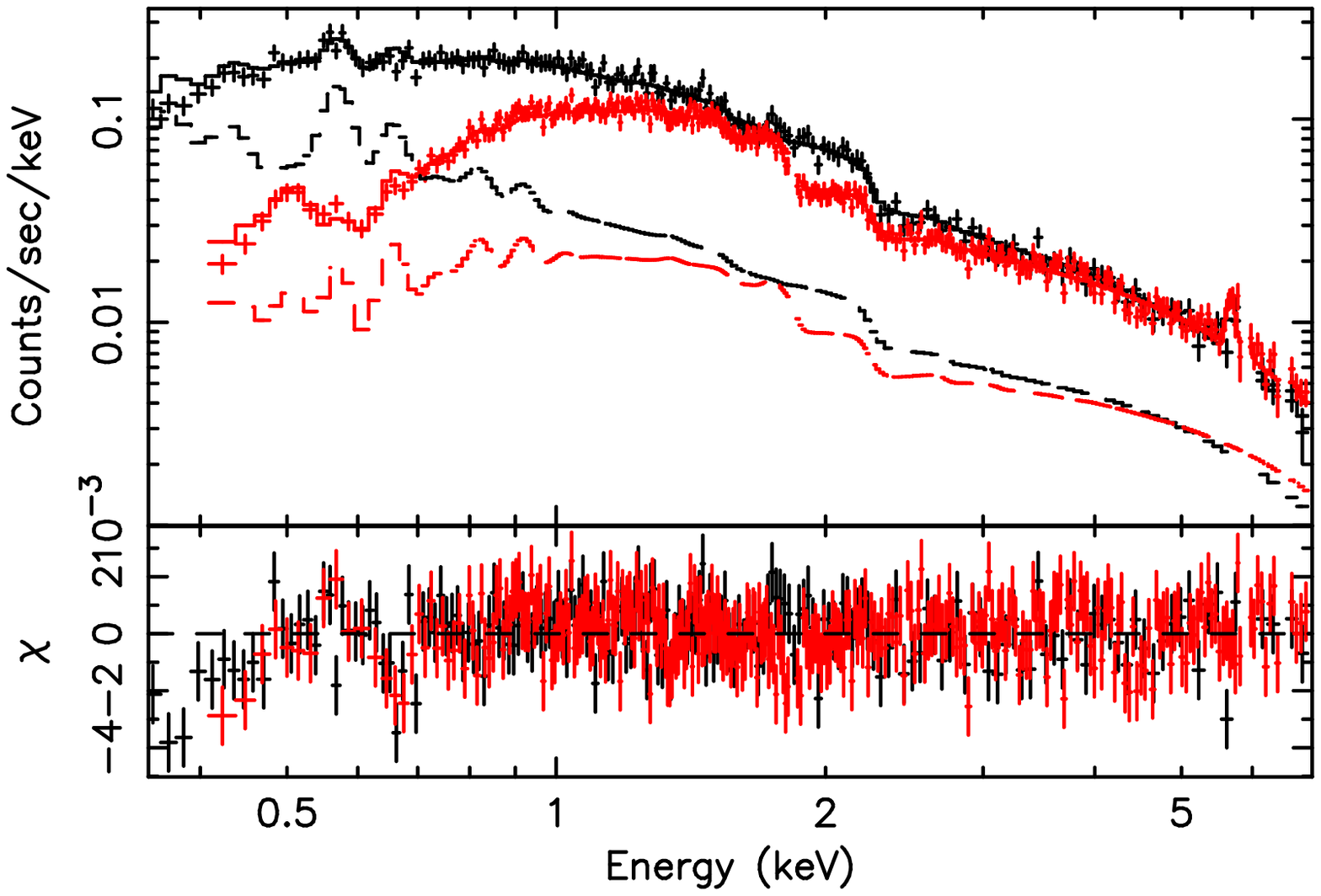}
\end{center}
\caption{
Spectra fitted with the best-fit models, and residuals of A2218
observations, in which the data from the two observations were merged.
Galactic components are fixed to (left) Offset-A, and (right) Offset-B
models, respectively.
Black and red indicate BI and FI sensors.
The contribution of the Galactic emission, determined by offset
 observations, are also indicated by dashed lines.
}
\label{fig:a2218-offsetfixed}
\end{figure}

The A2218 spectrum for the radial region of $3'-8'$ was analyzed
here. The summed FI and the BI spectra were fitted
simultaneously. Before testing the warm emission in the model, we
first fitted with a model only including the thin thermal plasma (ICM)
and the background emission determined from the offset-region spectra.
We tried two models for the background: namely, the best-fit models
for the BI spectra determined from the Offset-A and Offset-B
observations.  We fixed all the parameters for the background spectrum
consisting of LHB, MWH and CXB at the best-fit values. The redshift of
the hot ICM was fixed at 0.1756.  We allowed the FI/BI normalization
to be free again.  
The results of the fit are summarized in
Table~\ref{tab:a2218-offsetfixed} and in
Fig.~\ref{fig:a2218-offsetfixed}.  
The different background (Offset-A or B) gives an ICM temperature
differing by about 10\%,
while similar metal abundance values were obtained in the two
cases.
The temperature and abundance are
generally consistent with previously determined values; {\it
XMM-Newton} spectra indicates $kT\sim 5~\mathrm{keV}$ in $3'-5'$ and
$Z=0.13\pm0.04$~solar in the central $5'.1$ region
\citep{2005A&A...433..777P}, while {\it Chandra} estimated $kT$ to be
$6.9\pm0.5$ and $Z$ to be $0.20\pm0.13$ in the central $5'.1$ region
\citep{2002ApJ...567..188M}.  
The detailed analysis of the temperature and abundance of the
ICM will be reported elsewhere.

The spectra were well fitted with
this simple model, suggesting that no obvious soft excess is
present. 
The residual of the spectral fit should indicate a feature, such as
that caused by redshifted oxygen lines, if warm emission indeed gives a
significant contribution. Both panels in Fig. 5 suggest some features
around 0.57 keV and 0.68 keV: the former is a positive
residual, while the latter negative. 
Both features are near the strong  Galactic O emission lines:
O\emissiontype{VII} at 574 eV and O\emissiontype{VIII} at
654 eV, and not identical to
those of redshifted O lines at the redshift of A2218.
Their levels are less than that of Galactic emission
by a factor of 3--10.
Hence, they may originate in the fluctuation or incomplete
modeling of  Galactic emission.
Note that the
O\emissiontype{VIII} line (654 eV) at the source
is redshifted to 556 eV, which is close to the strong 
O\emissiontype{VII}  line (574 eV) in the Galactic emission.
Thus it is rather difficult to detect the
warm gas with this line. The redshifted 
O\emissiontype{VII} line (574 eV shifts to
488 eV) is also close to the peak caused by the oxygen edge of the
detector deadlayer and OBF, in particular for the FI sensors
(see Fig.~4 of Koyama et~al. 2006).
These situations make the XIS sensitivity to warm gas 
for a source with redshift similar to that of A2218
poorer than it is at other redshifts. 
We will carry out a quantitative evaluation in
the next section.

\begin{table}[hbt]
\begin{center}
\caption{Best-fit parameters of A2218 with single temperature model
\label{tab:a2218-offsetfixed}
}
{\normalsize
\begin{tabular}{lcc}
\hline \hline
Parameter & Offset-A Case & Offset-B Case\\
\hline 
LHB $kT$ (keV) & 0.08 (fixed) & 0.08 (fixed) \\
LHB Normalization$^\mathrm{a}$
 & $3.2\times10^{-6}$ (fixed)
 & $6.2 \times10^{-6}$  (fixed)\\
MWH $kT$ (keV) & $0.158$ (fixed) & $0.248$ (fixed)\\ 
MWH Normalization$^\mathrm{a}$ & $6.7\times10^{-7}$ (fixed)
  & $5.0\times10^{-7}$ (fixed)\\
CXB Photon index & 1.4 (fixed) & 1.4 (fixed)\\
CXB Normalization$^\mathrm{b}$ &  $8.5\times10^{-7}$ (fixed)
 &  $8.5\times10^{-7}$ (fixed)\\
\hline
ICM $kT$ (keV) & $5.40^{+0.27}_{-0.15}$ & $6.00\pm0.22$\\
ICM $Z$ (solar) & $0.20\pm0.04$ & $0.21\pm 0.04$\\
ICM $z$ & 0.1756 (fixed) & 0.1756 (fixed) \\
ICM Normalization$^\mathrm{a}$ & $2.08\pm0.03\times10^{-5}$
 & $2.00\pm0.03\times10^{-5}$ \\
 FI/BI ratio & $0.89\pm0.01$ & $0.90\pm0.01$\\
\hline
$\chi^2/\mathit{dof}$ & 475.21/475 & 520.72/475 \\
\hline
\multicolumn{3}{l}{ \parbox{10cm}
{$^\mathrm{a}$~
$\int n_\mathrm{e}n_\mathrm{H}~dV/4\pi(D_\mathrm{A}(1+z))^2$ per
solid angle in units of $10^{14}~ \mathrm{cm^{-5}~ arcmin^{-2}}$,
where $n_\mathrm{e}$ is the electron density,
$n_\mathrm{H}$ the hydrogen density, and $D_\mathrm{A}$ 
the angular size distance.
}
}\\
\multicolumn{3}{l}{$^\mathrm{b}$~
 In units of $\mathrm{photons~cm^{-2}~s^{-1}~keV^{-1}~arcmin^{-2}}$ at 1 keV}
\end{tabular}
}
\end{center}
\end{table}

\subsubsection{Constraint on redshifted O lines}

We performed another spectral fit by adding two Gaussian emission lines
to the model, in order to test for the existence of redshifted O lines.
The energies of the lines were fixed to 488.22 eV and 555.99 eV, which
correspond to O\emissiontype{VII} (resonance)
and O\emissiontype{VIII} lines at
$z=0.1756$.  
Although the energy of the O\emissiontype{VII} line could be at
most 10~eV lower due to the contribution of intercombination and
forbidden lines, we confirmed that
the results shown below are not affected by that
difference in energy.
The intrinsic widths were fixed to zero.  The temperature
and abundance of the ICM component were free parameters, while all the
background parameters (for LHB, MWH and CXB) 
including the normalizations
were fixed at the
best-fit values determined with the
Offset data described in \S~\ref{sec:Single-temp-model}.  The free
parameters were the temperature $kT$, abundance $Z$ and normalization
of the ICM, the surface brightness $I$ of redshifted
O\emissiontype{VII} and O\emissiontype{VIII} lines, and FI/BI
normalization ratio.  We again tried the two background models,
determined from the Offset-A and Offset-B observations, respectively.

\begin{figure}[hbt]
\begin{center}
\FigureFile(0.47\columnwidth,clip){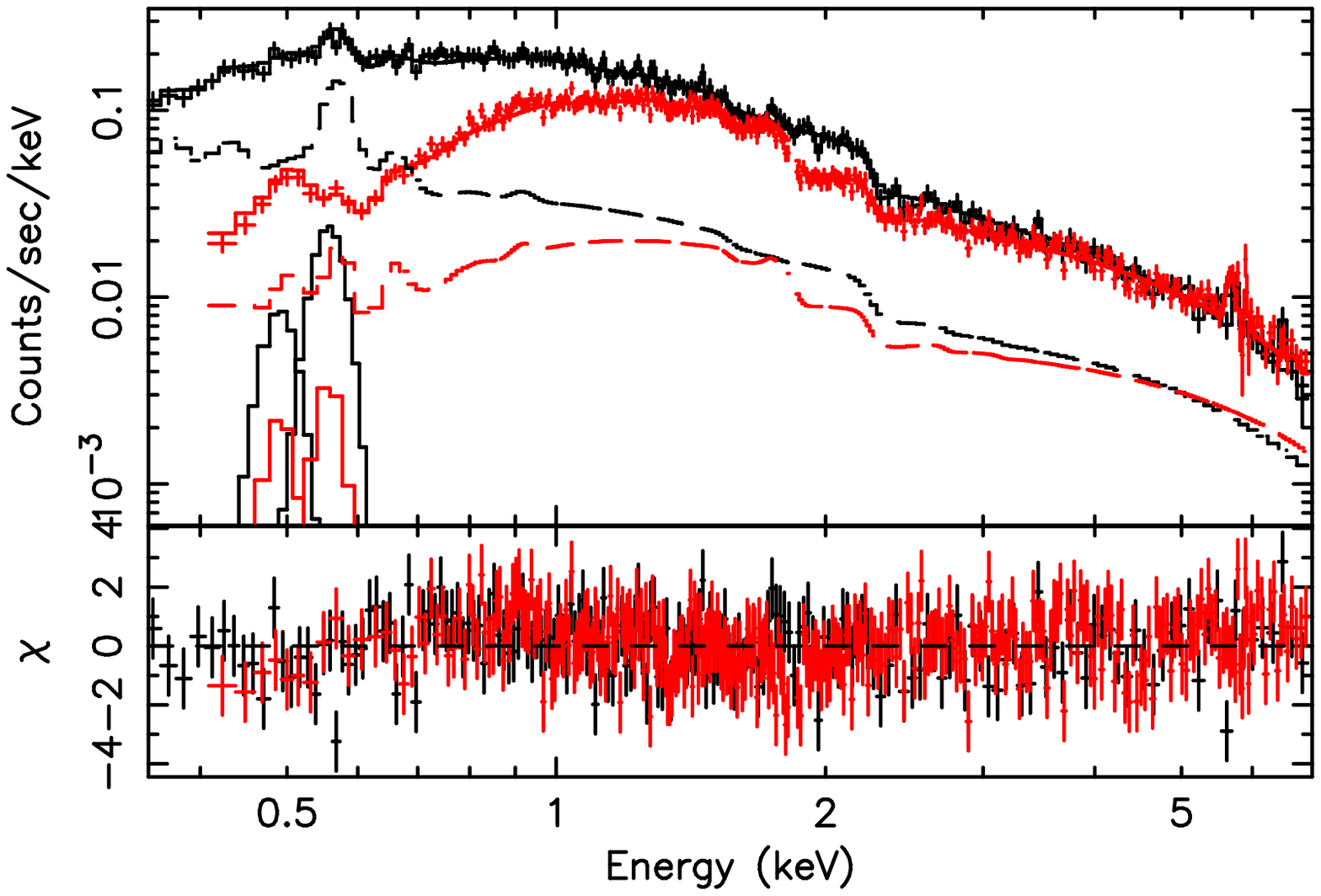}
\hspace{0.03\columnwidth}
\FigureFile(0.47\columnwidth,clip){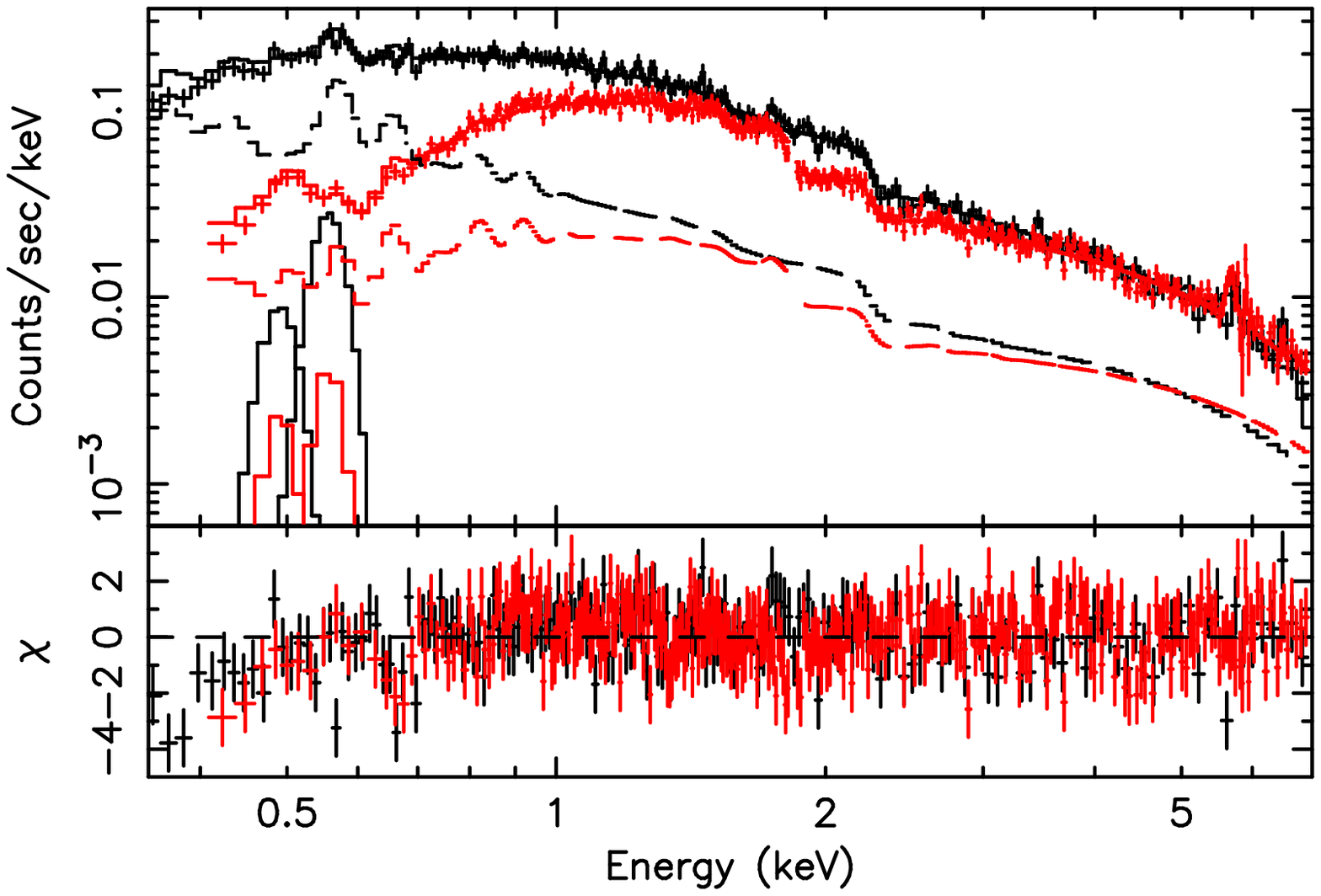}
\\ \vspace{0.5cm}
\FigureFile(0.47\columnwidth,clip){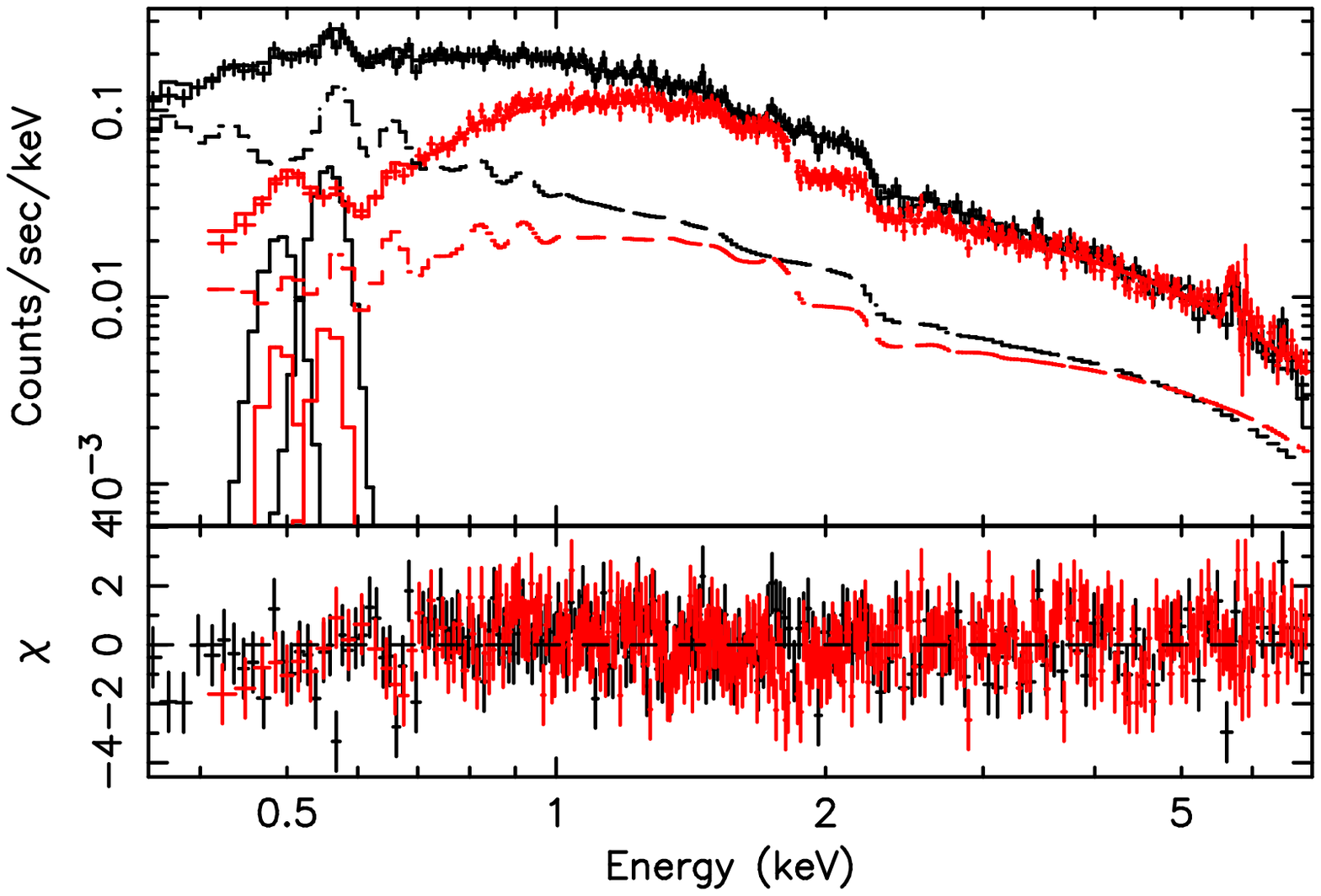}
\end{center}
\caption{A2218 BI (black) and FI (red) spectra
and models with redshifted O lines.
The two emission lines indicated with solid lines in the 0.4--0.6~keV range 
show the maximum allowed level (2$\sigma$)
of O\emissiontype{VII} and O\emissiontype{VIII} lines.
Dashed lines indicate the background emission.
(upper left) without considering systematic uncertainty, using the
best-fit model of Offset-A observation as the Galactic emission.
(upper right) without considering systematic uncertainty, using Offset-B
observation.
(bottom) the model of maximum O line intensity after
considering systematic uncertainties: with
ARFs of 20\% thicker contaminant and
10\% fainter Galactic model than the one determined from the Offset-B
observation. The O\emissiontype{VII} and
O\emissiontype{VIII} surface brightness was
$1.1\times10^{-7}$% ~$\mathrm{photons~cm^{-2}~s^{-1}~arcmin^{-2}}$
and 
$2.7\times10^{-7}$~$\mathrm{photons~cm^{-2}~s^{-1}~arcmin^{-2}}$,
respectively.
}
\label{fig:modelwithOupperlimits}
\end{figure}

The best-fit parameters and improvement in $\chi^2$ values are shown
in Table~\ref{tab:a2218-offsetfixed-gauss}.  The obtained surface
brightness $I$ of the lines is small and consistent with zero;
the improvement in $\chi^2$ is less than 2.71 (90\% significance
for 1 free parameter).  We then constrained the allowed O line
intensity range.  The upper limits for the O\emissiontype{VII} and
O\emissiontype{VIII} $I$ were estimated from that causing an increment
of the $\chi^2$ value by 4.0 ($2\sigma$ limit) over the minimum
(best-fit) level.  In this process, we allowed the other free
parameters (i.e., the ICM temperature, abundance, normalization, and
FI/BI ratio) to vary. The upper limits of the two lines were
determined separately, and the values are also shown in
Table~\ref{tab:a2218-offsetfixed-gauss} in parentheses.
Fig.~\ref{fig:modelwithOupperlimits} shows the model with the
upper-limit intensities of the O\emissiontype{VII} and
O\emissiontype{VIII} lines in the upper 2 panels; the left panel is
the case with the background of Offset-A, while the right panel is for
Offset-B.  The upper limits are shown with solid lines,
while the background spectrum is shown with dashed lines.  The upper
limits of the two O lines are 10--20\% of the background
level at the center energy of each line.

It should be noted that the O\emissiontype{VIII} line is
also produced by the ICM ($T\sim 5$~keV), 
while O\emissiontype{VII} is not. 
The intensity, which depends on the temperature and O abundance,
is $2\times10^{-8}~\mathrm{photons~ cm^{-2}~ s^{-1}~ arcmin^{-2}}$
assuming the parameters in Table~\ref{tab:a2218-offsetfixed}.
However, this value contains an uncertainty; 
while the abundance was mostly determined by the Fe K and L features
in the fit, Fe and O abundances may have a different value. 
We estimated the upper limit of $I$ for the 
redshifted O\emissiontype{VIII} 
line from the WHIM in the extreme case of no O in the ICM, 
because this situation gives the largest contribution from the WHIM. 
The upper limit then
increased to $1.6\times10^{-7}~\mathrm{photons~ cm^{-2}~
s^{-1}~ arcmin^{-2}}$ and
$1.8\times10^{-7}~\mathrm{photons~ cm^{-2}~ s^{-1}~ arcmin^{-2}}$
for the Offset-A and Offset-B backgrounds, respectively.

Next, we investigated how the upper limits of $I$
change by considering the
systematic uncertainties described in \S~\ref{sec:Analysis-method}.
We already checked one of these uncertainties in
Table~\ref{tab:a2218-offsetfixed-gauss}: variation of the Galactic
emission between Offset-A and Offset-B fields.  The difference is
$\lesssim 20\%$.  When we used the ARFs with 20\% thicker contaminant,
the upper limits of the O\emissiontype{VII} and O\emissiontype{VIII}
lines increased by $\sim130\%$ and $\sim30\%$, respectively.  These
limits further increase by $\sim15\%$ and $\sim30\%$ for the
respective lines, if we adopt 10\% fainter Galactic emission.  On the
other hand, the upper limits do not increase when we select the data
for the low background condition (COR$>8~\mathrm{GeV}~c^{-1}$)
or when the energy resolution of the detector was artificially degraded
by 5--35~eV.
To summarize, the
conservative upper limits are derived by assuming no O in the ICM
and employing the ARFs with 20\%
thicker contaminant and 10\% fainter Galactic model for the Offset-B
observation; the values are $1.1\times10^{-7}$ 
~$\mathrm{photons~cm^{-2}~s^{-1}~arcmin^{-2}}$ and
$3.0\times10^{-7}$~$\mathrm{photons~cm^{-2}~s^{-1}~arcmin^{-2}}$ for
O\emissiontype{VII} and O\emissiontype{VIII} lines, respectively.  We
adopt these values as the upper limits of the O line intensity around 
A2218, considering both statistical and systematic errors.  The
spectra with the model with these upper limits are shown in the lower
panel of Fig.~\ref{fig:modelwithOupperlimits}. Note that even after
considering these systematic errors, the upper limits are still $\sim
30\%$ of the level of the Galactic emission at the same energy.

%%%
\begin{table}[hbt]
\begin{center}
\caption{Best-fit parameters using a model with redshifted O lines
\label{tab:a2218-offsetfixed-gauss}
}
{\normalsize
\begin{tabular}{lcc}
\hline \hline
Parameter & Offset-A Case & Offset-B Case\\
\hline
ICM $kT$ (keV) & $5.41^{+0.28}_{-0.16}$ & $6.02\pm0.22$\\
ICM $Z$ (solar) & $0.20\pm0.04$ & $0.21\pm0.04$\\
ICM $z$ & 0.1756 (fixed) & 0.1756 (fixed) \\
ICM Normalization$^\mathrm{a}$ & $2.08\pm0.03\times10^{-5}$
 & $2.00\pm0.03\times10^{-5}$ \\
% \hline 
O\emissiontype{VII} $E$ (eV) & 488.22 (fixed) & 488.22 (fixed) \\
O\emissiontype{VII} $I^\mathrm{b~c}$ 
  & 0 ($<3.7\times10^{-8}$) & 0 ($<3.9\times10^{-8}$) \\
O\emissiontype{VIIII} $E$ (eV) & 555.99 (fixed) & 555.99 (fixed) \\
O\emissiontype{VIII} $I^\mathrm{b~c}$ 
  & $4.0~(<13.4)\times10^{-8}$ 
 & $6.8~(<15.8)\times10^{-8}$ \\
FI/BI ratio & $0.89\pm0.01$ & $0.90\pm0.01$\\
\hline
$\Delta\chi^2/\Delta\mathit{dof}$ & 0.78/2 & 2.25/2 \\
\hline
\multicolumn{3}{l}{ \parbox{10cm}
{$^\mathrm{a}$~
$\int n_\mathrm{e}n_\mathrm{H}~dV/4\pi(D_\mathrm{A}(1+z))^2$ per
solid angle in units of $10^{14}~ \mathrm{cm^{-5}~ arcmin^{-2}}$,
where $n_\mathrm{e}$ is the electron density,
$n_\mathrm{H}$ the hydrogen density, and $D_\mathrm{A}$ 
the angular size distance.
}
}\\
\multicolumn{3}{l}{$^\mathrm{b}$~
  In units of $\mathrm{photons~cm^{-2}~s^{-1}~arcmin^{-2}}$
} \\
\multicolumn{3}{l}{$^\mathrm{c}$~
 Upper limits are quoted in 2$\sigma$ confidence level.
}
\end{tabular}
}
\end{center}
\end{table}

\section{Discussion}

We have observed A2218 with the XIS instrument onboard {\it Suzaku} to
search for the redshifted O emission lines from the WHIM, which is
possibly forming a large-scale filament around the cluster.  The
object was selected because its redshift ($z=0.1756$) would allow the XIS
to separate the WHIM lines from the Galactic ($z=0$) ones, and because
an elongated structure in the line-of-sight direction was suggested
from the previous studies of this cluster.  We detected no
redshifted O lines, and set a constraint on their intensities.  In this
section, we compare our constraints on the line intensity with those
reported in other works and also discuss the future prospect of
studying the WHIM emission with the {\it Suzaku} XIS.

\subsection{Comparison with other results}

\begin{figure}[hbt]
\begin{center}
\FigureFile(0.47\columnwidth,clip){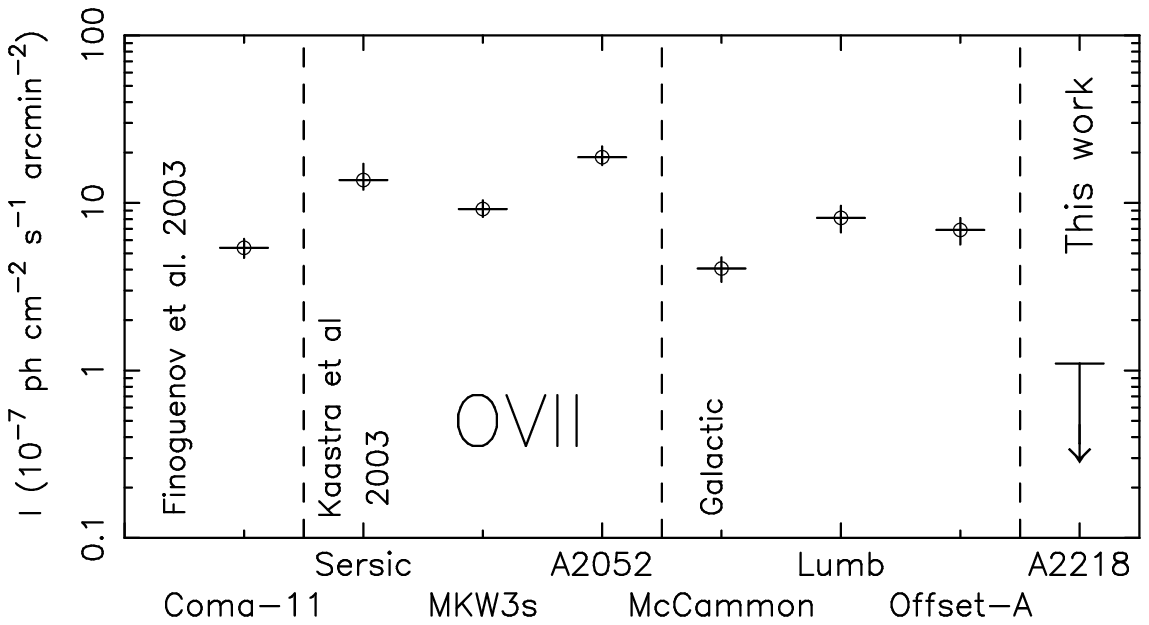}
\hspace{0.03\columnwidth}
\FigureFile(0.47\columnwidth,clip){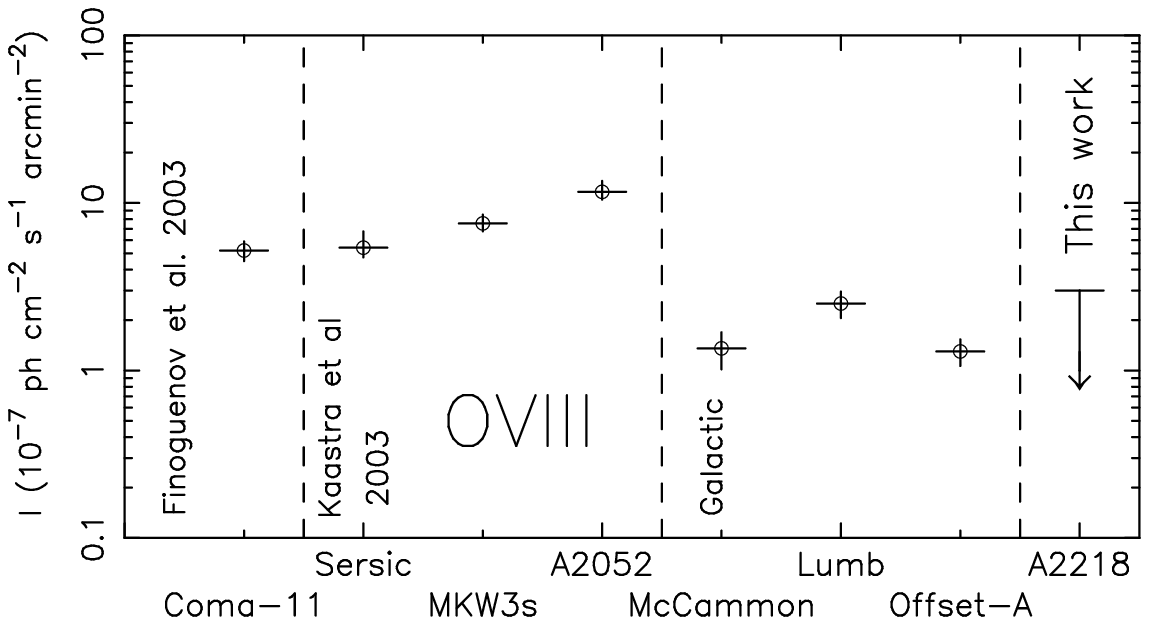}
\end{center}
\caption{Comparison of 
O\emissiontype{VII} (left) and O\emissiontype{VIII} (right) surface
brightness.
From left to right, those in the Coma-11 field
\citep{finoguenov03:_xmm_newton_x_coma},
S\'{e}rsic~159--03, MKW 3s and A2052
\citep{2003A&A...397..445K},
Galactic emission of 
\cite{2002ApJ...576..188M},
\citet{2002A&A...389...93L},
and our Offset-A observation,
and the upper limits in A2218 outskirts (this work) are plotted.
}
\label{fig:compareoxygen}
\end{figure}

\citet{2003A&A...397..445K} and \citet{finoguenov03:_xmm_newton_x_coma}
reported positive detections of O lines around clusters of galaxies
based on {\it XMM-Newton} observations.
\citet{2003A&A...397..445K} detected significant
O lines in three clusters, 
S\'{e}rsic~159--03, MKW 3s and A2052, 
and possible, but not uniquely proven, O lines in
the Coma and A1795 clusters.
\citet{finoguenov03:_xmm_newton_x_coma} detected the O lines in 
the outskirts of the Coma cluster, in particular,
the Coma-11 field.
Fig.~\ref{fig:compareoxygen} compares the O line surface brightness
$I$ for their observations and our results.  The left and right panels
are for the O\emissiontype{VII} and O\emissiontype{VIII} cases,
respectively.  The intensity of the O lines of Galactic emission
measured by a microcalorimeter experiment
of \cite{2002ApJ...576..188M},
the one compiled by
\citet{2002A&A...389...93L} and the one estimated from our Offset-A
observation are also shown.  The surface brightness quoted for
\citet{2003A&A...397..445K} was calculated using the temperature and
emission measure in their Table 7 and the metal abundance from
Tables 4 and 5 of \citet{2004A&A...420..135T},
assuming the solar abundance ratio Fe to O
\citep[Fe/O number density ratio of 0.55;][]{anders89:_abund}.
The surface
brightness levels of the O\emissiontype{VII} or O\emissiontype{VIII}
emission lines that were reported so far from the {\it XMM-Newton}
 observations are
similar to or higher than that of the Galactic emission.  In contrast,
the upper limit of O\emissiontype{VII} line intensity
 in the A2218 vicinity
obtained in this work is about six times lower than the Galactic
level.  The upper limit for O\emissiontype{VIII} is also lower than
the level reported as a positive detection in other works. The tight
upper limit we obtained here demonstrates the good spectral capability
of the XIS, in particular below 0.5 keV\@.

\subsection{Constraint on the WHIM density}

If we assume that the O\emissiontype{VII} emission line is produced in
a cloud with uniform density and temperature under collisional
ionization equilibrium, the surface brightness $I$ is determined by
the electron density $n_\mathrm{e}$, 
the hydrogen density $n_\mathrm{H}$, 
path length $L$, and metal
abundance $Z$ of the cloud as
\begin{equation}
 I = C(T)~ (1+z)^{-3}~ n_\mathrm{e}n_\mathrm{H}~ ZL,
\end{equation}
where $C(T)$ is a coefficient that depends on the temperature of the
cloud.  Here, we will constrain the density of the cloud assuming the
temperature to be $T\sim 2\times 10^{6}$~K, since the
O\emissiontype{VII} line is strongest at this temperature.
Substituting $C(T=2\times 10^{6}~\mathrm{K})$ with the value 
calculated with the SPEX code
\citep{1996uxsa.coll..411K}, our constraint, $I < 1.1 \times 10^{-7} $~$
\mathrm{photons~cm^{-2}~s^{-1}~arcmin^{-2}}$ at $z=0.1756$, 
and electron-to-hydrogen number density
ratio $n_\mathrm{e}/n_\mathrm{H}=1.2$ for
ionized gas,
gives the
following condition
\begin{equation}
 n_\mathrm{H} < 7.8\times10^{-5}~\mathrm{cm^{-3}}~
  \left(\frac{Z}{0.1~Z_\mathrm{\odot}}\right)^{-1/2}~
  \left(\frac{L}{2~\mathrm{Mpc}}\right)^{-1/2}.
\end{equation}
Note that $L=2$~Mpc corresponds to $11'$, the average diameter of the
annular region where we extracted the spectra. 
The overdensity $\delta\equiv n_{\rm
H}/\bar{n}_{\rm H}$ of this cloud is

\begin{equation}
 \delta < 270~
  \left(\frac{Z}{0.1~Z_\mathrm{\odot}}\right)^{-1/2}~
  \left(\frac{L}{2~\mathrm{Mpc}}\right)^{-1/2},
\end{equation}
where $\bar{n}_{\rm H} = X\Omega_\mathrm{b}\rho_\mathrm{crit} (1+z)^3
/m_\mathrm{p} = 1.77\times10^{-7}~(1+z)^3~\mathrm{cm^{-3}}$ is the
mean hydrogen density in the universe, in which $X=0.71$ is the
hydrogen-to-baryon mass ratio, $\Omega_\mathrm{b}=0.0457$ is the
baryon density of the universe,
$\rho_\mathrm{crit}=9.21\times10^{-30}~\mathrm{g~cm^{-3}}$ is the
critical density of the universe, and $m_\mathrm{p}$ is the proton
mass.  Even though this level of overdensity is much higher than the
typical WHIM density ($\delta \sim 10$), it shows that Suzaku can
certainly detect the high-density end of the WHIM distribution that is
predicted to exist near clusters.

\subsection{Prospect for the WHIM observation with the XIS}

Although we have set a tight constraint on the intensity of redshifted
O lines, the inherent sensitivity of the XIS was presumably not
achieved, since the redshifted O\emissiontype{VIII} line fell almost
on the Galactic O\emissiontype{VII} line, while the redshifted
O\emissiontype{VII} line overlapped with the instrumental neutral O
edge.  Therefore, better sensitivity is expected for clusters with a
redshift such that the redshifted lines are free from the Galactic and
instrumental features.  For example, a redshift of $z\sim0.07$ or
$z>0.2$ is suitable for this purpose.  The former case is more
promising, because the surface brightness of high-$z$ clusters is
lower and because the contamination of bright ICM emission in the
center of the cluster makes the detection of warm-hot emission in the
outskirts difficult for distant clusters.  Note that the 
XIS is capable of
separating a redshift difference of $\Delta z\sim0.07$ at the O line
energy.

Because of the contamination on the XIS OBF, the effective area at the
redshifted O\emissiontype{VII} line had dropped by $\sim 25\%$ at the
time of the A2218 observation.  If there had been no contamination on
the OBF, we could have obtained more photons by this fraction.  This
contamination also caused a large systematic uncertainty corresponding
to about 5\% of the OVII line flux.  Since the thickness and chemical
composition of the contaminant will be better understood as more data
are accumulated, the uncertainty will be less influential in future
observations.  Furthermore, we hope that the original XIS sensitivity
in the soft X-ray energy range can be recovered in the future by 
hardware changes
including warming the contaminated OBFs.

It should be noted that in most cases concerning the search for faint
WHIM emission with CCD sensors, the variation of Galactic emission
causes a larger uncertainty than the statistical one.  Offset
observations, such as those carried out in our work, are always
desirable to determine the background emission level reliably.
Looking at a somewhat longer time frame, 
observations with microcalorimeters are
promising, because their outstanding energy resolution ($\Delta E
\lesssim 7$ eV) for extended sources can easily separate the WHIM
emission from the Galactic spectral features.  
For example, two orders of magnitude
higher sensitivity than the present work is expected with a small
mission such as {\it DIOS}
\citep[Diffuse Ionized Oxygen Surveyor;][]{2007_ohashi_dios_spie}, 
which is proposed to be launched in early 2010s
\citep{2004PASJ...56..939Y}.

\section{Summary}
We used {\it Suzaku} XIS observations to constrain the intensity of
O emission lines around the cluster of galaxies A2218\@.  After
considering systematic uncertainties mainly caused by the Galactic
emission and the uncertainty in the detector response, we obtained
upper limits for the surface brightness of the O\emissiontype{VII} and
O\emissiontype{VIII} lines of $1.1\times10^{-7}$ 
~$\mathrm{photons~cm^{-2}~s^{-1}~arcmin^{-2}}$ and
$3.0\times10^{-7}$~$\mathrm{photons~cm^{-2}~s^{-1}~arcmin^{-2}}$,
respectively.  These upper limits are significantly lower than the
previously reported fluxes around other clusters of galaxies.  Our
tight constraints demonstrate the sensitivity of the XIS for redshifted O
lines.

\bigskip
We are grateful to Dr.\ S. Sasaki for valuable comments on
cosmological parameters 
and the anonymous referee for important comments.
YT would like to thank Prof. D. McCammon for
a lot of information and comments regarding the Galactic emission,
and Dr.\ N. White for careful internal review of the manuscript.
The authors would like to thank ISAS/JAXA and the entire {\it
Suzaku} team for the opportunity to participate in the development of
the {\it Suzaku} mission as members of its Science Working Group.  YT
is supported by grants from the JSPS Research Fellowship for Young
Scientists (DC 16-10681 and PD 18-7728).  This work is supported by
Grant-in-Aid for Science Research of JSPS and MEXT
(KAKENHI 14079103, 15340088 and 16002004), 
and NASA grant (NNG05GP87G, NNG05GM92G, NNG06GC04G).

%%%%%%%%%%%%%%%%%%%
% bibliography
%%% \bibliographystyle{apj}
%%% \bibliography{myrefs_en}

\end{document}